\documentclass[a4paper,byrevtex]{revtex4}
\usepackage{dcolumn}
\usepackage{amsmath}
\usepackage{bm}
\usepackage{graphicx}

\begin{document}

\title{Quantum and classical probability distributions for arbitrary Hamiltonian}

\author{Claude \surname{Semay}}
\email[E-mail: ]{claude.semay@umons.ac.be}
\author{Ludovic \surname{Ducobu}}
\email[E-mail: ]{ludovic.ducobu@student.umons.ac.be}
\affiliation{Service de Physique Nucl\'{e}aire et Subnucl\'{e}aire,
Universit\'{e} de Mons,
UMONS Research Institute for Complex Systems,
Place du Parc 20, 7000 Mons, Belgium}

\date{\today}

\begin{abstract}
In the limit of large quantum excitations, the classical and quantum probability distributions for a Schr\"odinger equation can be compared by using the corresponding WKBJ solutions whose rapid oscillations are averaged. This result is extended for one-dimensional Hamiltonians with a non-usual kinetic part. The validity of the approach is tested with a Hamiltonian containing a relativistic kinetic energy operator. 
\end{abstract}



\maketitle

\section{Introduction}
\label{sec:intro}

By many aspects, quantum theory is a very strange theory with numerous non-intuitive predictions. Nevertheless, our familiar classical world is the result of quantum phenomena at the atomic and subatomic levels. So, it is interesting to establish connections between classical and quantum descriptions, and to understand how the macroscopic world emerges from the microscopic world. An interesting approach for stationary quantum states is to compare the probability density given by the square modulus of the wave-function with a ``classical probability distribution" obtained from the corresponding classical equations of motion. It can then be shown that both functions approach each other, in the limit of large quantum excitations, once the rapid oscillations of the quantum density are averaged. The classical probability distribution can be compared directly with the explicit (analytical or numerical) corresponding quantum distribution for some particular Hamiltonians. This is done, for instance, in \cite{robi95,yode06} for one-dimensional Schr\"odinger equations. But a more general procedure is available. 

The WKBJ method, named after Wentzel, Kramers, Brillouin and Jeffreys \cite{flug99,maha09}, yields a semi-classical solution of a quantum problem, also in the limit of large quantum excitations. So it is possible to compare the  classical probability distribution directly with the averaged WKBJ solution for Schr\"odinger equations \cite{sen04,sen06}. In this paper, the same approach is generalized for one-dimensional Hamiltonians with an arbitrary kinetic energy. Such Hamiltonians are used in several domains: atomic physics with non-parabolic dispersion relation \cite{arie92}, hadronic physics with particle masses depending on the relative momentum \cite{szcz96}, quantum mechanics with a minimal length \cite{brau99,ques10}.

The characteristics of a general Hamiltonian with a non-usual kinetic energy are given in Sec.~\ref{sec:hamiltonian}, where natural constraints are given on the kinetic part. The notion of classical probability distribution for the usual Schr\"odinger equation is recalled in Sec.~\ref{sec:classical}, and extended to the case of more general Hamiltonians. In sec.~\ref{sec:wkbj}, the WKBJ approximation is generalized for this type of Hamiltonians, and the connection is made between the classical probability distribution and the quantum probability distribution obtained from the WKBJ method. Some examples are treated in Sec.~\ref{sec:examples}, and concluding remarks are given in Sec.~\ref{sec:conclusion}.

\section{The Hamiltonian}
\label{sec:hamiltonian}

The following general one-dimensional Hamiltonian is considered
\begin{equation}
\label{TpVx}
H=T(p)+V(x),
\end{equation}
where $T(p)$ is the kinetic part depending on the momentum $p$, and $V(x)$ the potential part depending on the position $x$. Variables $p$ and $x$ are conjugate: $[x,p]=i\hbar$. This Hamiltonian can correspond to a particle in a potential well if $x$ is interpreted as the distance from the origin, or to two particles in mutual interaction if $x$ is interpreted as the relative distance. It is assumed that bound states are supported by this Hamiltonian and that the potential well $V(x)$ has no singularity. Obviously, the form of the kinetic energy $T$ cannot be completely arbitrary. Four conditions are imposed:
\begin{description}
  \item{A.} $T(p) \ge 0$ for all values of the momentum $p$, in order that the kinetic energy is a positive quantity. This sounds physically reasonable, but this is not necessary from a mathematical point of view.
  \item{B.} $T(p)=T(-p)$. It seems reasonable that the kinetic energy is an even function of the momentum in order that it cannot be dependent on the direction of propagation of the particle. 
  \item{C.} $T(p)$ is a monotonically increasing function of $|p|$. It seems quite natural that the kinetic energy increases with the modulus of the momentum.
  \item{D.} $T(p)$ is at least of class $C^2$. The utility of this condition will appear below.
\end{description}

The speed of the particle is defined using the usual Hamilton's equations
\begin{equation}
\label{speed}
v(p) = \frac{\partial H}{\partial p}= \frac{\partial T}{\partial p} =T'(p).
\end{equation}
This is in agreement with the phenomenological definition given in \cite{arie92,sema13}. Moreover, $v(-p) = T'(-p) = -T'(p) = -v(p)$, since $T'(p)$ is an odd function because of condition~(B). So, to change the sign of the momentum is to change the sign of the speed, as expected. In particular, the speed is vanishing for a null momentum, $v(0) = T'(0) = 0$.

Using conditions (B) and (D) for small values of the momentum, one can write
\begin{equation}
\label{Texp}
T(p)= T(0)+ \frac{T''(0)}{2} p^2 + \textrm{O}(p^4).
\end{equation}
In this limit, Hamiltonian~(\ref{TpVx}) reduces to a usual Schr\"odinger Hamiltonian with an effective mass $M=1/T''(0)$ and a constant contribution $T(0)$ to the eigenenergies which can be identified with the rest energy of the particle. 

\section{The classical probability distribution}
\label{sec:classical}

The position of a particle can be, in principle, perfectly determined in a classical motion. So, for a classical probability distribution $\rho_\textrm{cl}(x)$ to make sense in this context, it is necessary to introduce a random procedure into the problem. For example, one can choose to perform a measurement of the position at a random time. As only bounded one-dimensional motions in a potential well are considered here, the motion is periodic, with a period $\tau$, and the particle bounces back and forth between two classical turning points (TP) at $x=a$ and $x=b$ ($b > a$). We can then define the classical probability $\rho_\textrm{cl}(x)\, dx$ as the probability to find the particle into the interval $[x,x+dx]$. This gives 
\begin{equation}
\label{rhox}
\rho_\textrm{cl}(x)\, dx = \frac{2}{\tau}\,  dt(x) = \frac{2}{\tau}\, \frac{dx}{|v(x)|},
\end{equation}
where $v(x)$ is the speed of the particle. The absolute value insures that the probability is a positive number (the measurement is blind to the sense of propagation). 
The distribution is correctly normalized since the motion from left to right is identical to the motion from right to left, and then
\begin{equation}
\label{normrhox}
\int_a^b \rho_\textrm{cl}(x)\, dx = \int_{t_a}^{t_b} \frac{2}{\tau}\,  dt = 1.
\end{equation}
Definition~(\ref{rhox}) for $\rho_\textrm{cl}(x)$ seems quite natural since a particle is more likely measured at positions where it travels slowly. In a classical motion, the particle cannot exist outside the two TP. So, $\rho_\textrm{cl}(x) = 0$ for $x < a$ or $x > b$.

For a stationary solution, the energy $E$ of the particle is constant and is given by
\begin{equation}
\label{ener}
E = T(p) + V(x).
\end{equation}
Thanks to conditions~(B) and (C), a function $T^{-1}(p)$ can be defined such that $T^{-1}(T(p))=p$ for $p\ge 0$. Using~(\ref{ener}), the speed modulus of the particle can be written as a function of $x$ by using the definition~(\ref{speed})
\begin{equation}
\label{vx}
|v(x)| = T'(T^{-1}(E-V(x))),
\end{equation}
with $E-V(x) \ge 0$ for the classical motion. The probability distribution is then given by
\begin{equation}
\label{rhovx}
\rho_\textrm{cl}(x) = \frac{2}{\tau}\, \frac{1}{T'(T^{-1}(E-V(x)))} .
\end{equation}
We consider situations for which only two TP exist. The speed of the particle vanishing at TP, they are solutions of the equation
\begin{equation}
\label{tp}
V(a)=V(b)=E-T(0)=E_B,
\end{equation}
where $E_B$ is the binding energy of the particle. The distribution $\rho_\textrm{cl}(x)$ diverges at TP, but this is not a problem, provided the normalization condition (\ref{normrhox}) is satisfied. 

The kinetic part $t(p)=T(p)-T(0)$ vanishes for a null momentum. It has the following properties: $t'(p)=T'(p)$ and $t^{-1}(y)=T^{-1}(y+T(0))$. So,
\begin{equation}
\label{tp0}
T'(T^{-1}(E-V(x)))=t'(t^{-1}(E_B-V(x))).
\end{equation}
This shows that the presence of a rest energy does not influence the dynamics of the system.

\section{The WKBJ approximation}
\label{sec:wkbj}

The WKBJ approximation relies on a semi-classical expansion of the wave-function $\psi(x)$ of the form 
\begin{equation}
\label{wkbj}
\psi(x)=\exp\left(\frac{i}{\alpha}\sigma(x)\right),
\end{equation}
with the parameter $\alpha\to 0$ \cite{flug99,maha09}. Usually, this parameter $\alpha$ is simply posed as $\hbar$. But it is more satisfactory, from a mathematical point of view, to build a dimensionless quantity depending on $\hbar$ and other characteristic parameters of the system under study. In \cite{sen04,sen06}, for non-relativistic systems ($t(p)=p^2/(2 m)$), it is suggested to take
\begin{equation}
\label{alpha}
\alpha = \frac{\hbar}{\sqrt{2 m\, |E_B|}\, d},
\end{equation}
where $d=b-a$ is the distance between the two TP, $m$ is the mass of the particle. The semi-classical limit can then be reached for high value of the energy $|E_B|$, large mass $m$ of the particle or large size of the classical region $d$. This parameter appears naturally in a dimensionless rewriting of the Schr\"odinger equation. With a non-standard kinematics, this parameter must be redefined since there is a priori no automatic equivalent of the parameter $m$. This will be done at the end of this section. In the following, it is simply assumed that $\alpha$ can always be determined.

The computation of the WKBJ approximation for a Schr\"odinger equation can be found in many textbooks. But with a non-standard kinetic part, the derivation is more involved. The procedure developed here is inspired from a calculation performed in \cite{gold60} for a non-relativistic WKBJ approximation computed in the momentum space. The equation to solve is
\begin{equation} 
\label{3.3}
T\left(-i\hbar \frac{d}{d x}\right) \psi(x) + V(x) \psi(x) = E \psi(x)
\end{equation}
where the kinetic operator is defined by its Taylor expansion
\begin{equation} 
\label{3.3a}
T\left(-i\hbar \frac{d}{d x}\right) = \sum_{n=0}^{\infty} \frac{T^{(n)}(0)}{n!} \left(-i\hbar \frac{d}{d x}\right)^{n}.
\end{equation}
Condition (D) is not sufficient to guarantee the relevance of this last expression. This will be commented below. For the moment, (\ref{3.3a}) is assumed correct. As the limit $\alpha\to 0$ will be considered, it can be shown by induction that
\begin{equation}
\label{prop3.1}
\left(-i\hbar \frac{d}{d x}\right)^n e^{\frac{i}{\alpha} \sigma(x)} = \epsilon^n\, e^{\frac{i}{\alpha}\sigma(x)} \left( \left(\frac{d \sigma}{d x}\right)^n - i \alpha \frac{n(n-1)}{2} \frac{d^2 \sigma}{d x^2} \left(\frac{d \sigma}{d x}\right)^{n-2} + \textrm{O}(\alpha^2)\right),
\end{equation}
where $\epsilon=\hbar/\alpha$. The combination of (\ref{3.3a}) and (\ref{prop3.1}) gives
\begin{equation}
T\left(-i\hbar \frac{d}{d x}\right) e^{\frac{i}{\alpha} \sigma(x)} = e^{\frac{i}{\alpha} \sigma(x)} \left[ T\left(\epsilon \frac{d \sigma}{d x}\right) - \frac{i \alpha}{2} \epsilon^2 \frac{d^2 \sigma}{d x^2} T''\left(\epsilon \frac{d \sigma}{d x}\right) + \textrm{O}(\alpha^2)\, T(\epsilon) \right].
\end{equation}
Putting this last result in (\ref{3.3}), and dropping the exponential factor, gives
\begin{equation}
\label{3.5}
T\left(\epsilon \frac{d \sigma}{d x}\right) - \frac{i \alpha}{2} \epsilon^2 \frac{d^2 \sigma}{d x^2} T''\left(\epsilon \frac{d \sigma}{d x}\right) + V(x) + \textrm{O}(\alpha^2) = E,
\end{equation}
where the coefficient $T(\epsilon)$ is reabsorbed in $\textrm{O}(\alpha^2)$. The function $\sigma(x)$ can also be expanded in powers of $\alpha$
\begin{equation}
\label{3.5b}
\sigma(x) = \sigma_0(x) + \alpha\, \sigma_1(x) + \textrm{O}(\alpha^2),
\end{equation}
where all function $\sigma_j(x)$ are assumed to be independent of $\alpha$. In this case, an obvious result is
\begin{equation}
\left(\frac{d \sigma_0}{d x} + \alpha \frac{d \sigma_1}{d x}\right)^n  = \left(\frac{d \sigma_0}{d x}\right)^n + n \left(\frac{d \sigma_0}{d x}\right)^{n-1} \alpha \frac{d \sigma_1}{d x} + \textrm{O}(\alpha^2),
\end{equation}
The last equation yields
\begin{eqnarray}
T\left(\epsilon \frac{d \sigma}{d x}\right) & =& \sum_{n=0}^{\infty} \frac{T^{(n)}(0)}{n!} \epsilon^n \left(\frac{d \sigma_0}{d x} + \alpha \frac{d \sigma_1}{d x}\right)^n + \textrm{O}(\alpha^2) \nonumber \\ 
& =& T\left(\epsilon \frac{d \sigma_0}{d x}\right) + \alpha\, \epsilon \frac{d \sigma_1}{d x} 
T'\left(\epsilon \frac{d \sigma_0}{d x}\right) + \textrm{O}(\alpha^2)
\end{eqnarray}
and
\begin{eqnarray}
\alpha\, T''\left(\epsilon \frac{d \sigma}{d x}\right) & =& \alpha\, \sum_{n=0}^{\infty} \frac{T^{(n)}(0)}{n!} n (n-1) \left(\epsilon \frac{d \sigma_0}{d x}\right)^{n-2} + \textrm{O}(\alpha^2) \nonumber\\ 
& =& \alpha\, T''\left(\epsilon \frac{d \sigma_0}{d x}\right) + \textrm{O}(\alpha^2).
\end{eqnarray}
Finally, (\ref{3.5}) can be written
\begin{eqnarray}
\label{3.6}
&& \left[T\left(\epsilon \frac{d \sigma_0}{d x}\right) + V(x) - E\right] \nonumber \\ 
&& + \alpha \left[\epsilon \frac{d \sigma_1}{d x} T'\left(\epsilon \frac{d \sigma_0}{d x}\right) - \frac{i}{2}
\epsilon^2 \frac{d^2 \sigma_0}{d x^2} T''\left(\epsilon \frac{d \sigma_0}{d x}\right)\right] + \textrm{O}(\alpha^2) = 0.
\end{eqnarray}
The coefficient of each power of $\alpha$ must be vanishing, that is to say
\begin{eqnarray}
\label{3.7a}
&&T\left(\epsilon \frac{d \sigma_0}{d x}\right) + V(x) - E = 0, \\ 
\label{3.7b}
&&\frac{d \sigma_1}{d x} T'\left(\epsilon \frac{d \sigma_0}{d x}\right) - \frac{i}{2} \epsilon \frac{d^2 \sigma_0}{d x^2} T''\left(\epsilon \frac{d \sigma_0}{d x}\right) = 0.
\end{eqnarray}
Only a solution between the two TP is searched for. With the definition given above for the function $T^{-1}$, (\ref{3.7a}) implies
\begin{equation}
\label{epst}
\epsilon \frac{d \sigma_0}{d x} = \pm T^{-1}\left(E - V(x)\right),
\end{equation}
whose solution is given by
\begin{equation}
\label{3.8}
\sigma_0(x) = \pm \frac{1}{\epsilon} \int_x T^{-1}\left(E - V(y)\right) dy.
\end{equation}
The notation $\int_x\, f(y)\, dy$ denotes the integral of the function $f(y)$ with one limit at $x$ and the other limit at one of the TP. Equation~(\ref{3.7b}) can be written
\begin{equation}
\frac{d \sigma_1}{d x} = \frac{i}{2} \frac{d}{d x} \ln\left(T'\left(\epsilon \frac{d \sigma_0}{d x}\right)\right).
\end{equation}
Using (\ref{epst}), the solution of this last equation is given by
\begin{equation}
\sigma_1(x) = \frac{i}{2} \ln\left(T'\left(T^{-1}(E - V(x))\right)\right) + C,
\end{equation}
where $C$ is a constant. The argument of the $\ln$-function is positive thanks to the definition of the function $T^{-1}$. Finally, using~(\ref{wkbj}), the wave-function can be written
\begin{eqnarray}
\label{3.10}
\psi_{\textrm{WKBJ}}(x) &= & \frac{C_1}{\sqrt{T'\left(T^{-1}(E - V(x))\right)}} e^{+\frac{i}{\hbar}  \int_x T^{-1}\left(E - V(y)\right) dy} \nonumber \\ 
& +& \frac{C_2}{\sqrt{T'\left(T^{-1}\left(E - V(x)\right)\right)}} e^{-\frac{i}{\hbar} \int_x T^{-1}\left(E - V(y)\right) dy}, 
\end{eqnarray}
where $C_1$ and $C_2$ are normalization constants. The parameter $\alpha$ is not explicitly present is this expression. But, this approximation is only valid when $\alpha \ll 1$. The usual form is recovered for a non-relativistic kinematics.

In the case of bound states, the wave-function decays exponentially outside the TP. Inside, a calculation similar to the non-relativistic one \cite{flug99,maha09} gives
\begin{equation}
\label{3.14}
\psi_{\textrm{WKBJ}}(x)  = \frac{D}{\sqrt{T'\left(T^{-1}\left(E-V(x)\right)\right)}} \sin\left( \frac{1}{\hbar} \int_x^b T^{-1}\left(E-V(y)\right) dy + \beta \right).
\end{equation}
The normalization constant $D$ and phase angle $\beta$ are determined by matching this eigenfunction onto the evanescent wave-functions outside the TP. This procedure is not trivial because the WKBJ eigenfunction is a poor approximation to the actual eigenfunction near the TP. This problem has been solved by Langer in the non-relativistic case by using an explicit solution of the Schr\"odinger equation near the TP \cite{lang37}. In the case of a non-standard kinetic energy, the point must be reconsidered. But, very close to the TP, the particle is very slow and the kinetic energy can be replaced by the expansion~(\ref{Texp}). The general Hamiltonian~(\ref{TpVx}) reduces then to a Schr\"odinger Hamiltonian. Equation~(\ref{tp0}) shows that the presence of a rest energy does not change the dynamics. We can deduce that the result of Langer is still valid and that $\beta=\pi/4$.

Computations are performed in the position space with $p=-i\hbar\, d/dx$, but it is equivalent to work in the momentum space with $x=i\hbar\, d/dp$. There is no relevant differences between the results obtained by the two procedures, since the Fourier transform of the wave-function obtained by the WKBJ method in the position space is equal to the wave-function obtained by the WKBJ method in the momentum space up to a term $\textrm{O}(\alpha^2)$. This can be shown by a procedure which is similar to the one presented in \cite{gold60} for a non-relativistic kinematics. 

The quantification of the energy is obtained from the constraint that the wave-function~(\ref{3.14}) can be defined by integrating from one TP or from the other. The calculation is not different from the non-relativistic case \cite{flug99,maha09}, and the result is
\begin{equation}
\label{3.17}
\int_{a}^{b} T^{-1}\left(E-V(x)\right) dx = 
\int_{a}^{b} t^{-1}\left(E_B-V(x)\right) dx = \pi\,\hbar\left(n + \frac{1}{2}\right),
\end{equation}
where the quantum number $n$ is a positive integer including 0. The limits of integration depend on $E$ via relation~(\ref{tp}).

In principle, (\ref{3.3a}) demands a smooth behaviour of the kinetic operator, much more constraining that condition~(D). Nevertheless, once the WKBJ approximation is computed, it appears that non-smooth terms after the second derivative in the expansion of $T$ could probably spoil very little the quality of the approximation. This will be checked on an example in Sec.~\ref{sec:examples}. 

It is well known that the classical limit is reached for large values of the quantum number $n$, that is to say for large values of the excitation energy. In order to make apparent the role of $n$ in these cases, $E_B-V(x)$ is replaced by a constant, denoted here $E^*$. Then, (\ref{3.17}) reduces to 
\begin{equation}
\label{3.17b}
\int_{a}^{b} t^{-1}\left(E^*\right) dx \approx \pi\,\hbar \, n,
\end{equation}
that is to say
\begin{equation}
\label{3.17c}
\frac{1}{\pi\, n} \approx \frac{\hbar}{t^{-1}\left(E^*\right)\, d}.
\end{equation}
In the non-relativistic case, the right-hand side of (\ref{3.17c}) is the number $\alpha$ defined by (\ref{alpha}), provided $E^*$ is replaced by $|E_B|$. A natural definition of the parameter $\alpha$ for all types of kinematics is then
\begin{equation}
\label{alphans}
\alpha = \frac{\hbar}{t^{-1}\left(E^*\right)\, d} = \frac{\hbar}{T^{-1}\left(T(0)+E^*\right)\, d},
\end{equation}
where $E^*$ is an estimation of $E_B-V(x)$, and where $d$ depends also on $E_B$ by (\ref{tp}). If the potential has no singularity, $E^*$ can be chosen as $E_B-\min V(x)$ for instance. But an accurate computation of $E^*$ is not necessary to validate the method, since solutions~(\ref{3.14}) and (\ref{3.17}) do not explicitly depend on $\alpha$. It is just sufficient to be sure that a small parameter $\alpha$ can be defined for the system under study. The semi-classical regime is then reached when $\alpha \ll 1$, that is to say $n \gg 1$ since $\alpha\approx (\pi\, n)^{-1}$. Within these conditions, the variable argument of the sine function in (\ref{3.14}) can be approximated by 
\begin{equation}
\label{argsin}
\alpha^{-1} \frac{b-x}{d} \approx n\,\pi\frac{b-x}{d},
\end{equation}
and the sine function oscillates a great number of times between the two TP. 

The approximate quantum probability distribution is given by $\rho_{\textrm{WKBJ}}(x)=\psi_{\textrm{WKBJ}}^2(x)$ for $a < x < b$. For $x < a$ or $x > b$, it can be assumed that this distribution is vanishing since the wave-function decays exponentially. A classical approximation $\rho_{\textrm{cl WKBJ}}(x)$ for the quantum distribution $\rho_{\textrm{WKBJ}}(x)$ is obtained by replacing the rapidly oscillating square sine function by its average value $1/2$ \cite{sen04,sen06}. Finally, inside the two TP,
\begin{equation}
\label{rhoa}
\rho_{\textrm{cl WKBJ}}(x)=\frac{D^2}{2\,T'\left(T^{-1}\left(E-V(x)\right)\right)}.
\end{equation}
With proper normalizations, (\ref{rhovx}) and (\ref{rhoa}) are identical. This shows that, for a quite general one-dimensional Hamiltonian, the quantum probability distribution and the classical probability distribution approach each other, in the limit of large quantum excitations, once the rapid oscillations of the quantum density are averaged.

\section{Examples}
\label{sec:examples}

\begin{figure}[htb]
\includegraphics[width=5cm,height=3.27cm]{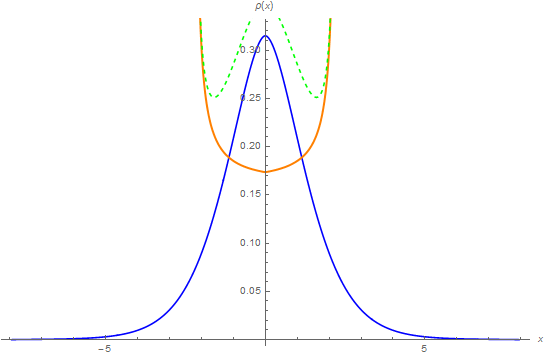}
\includegraphics[width=5cm,height=3.27cm]{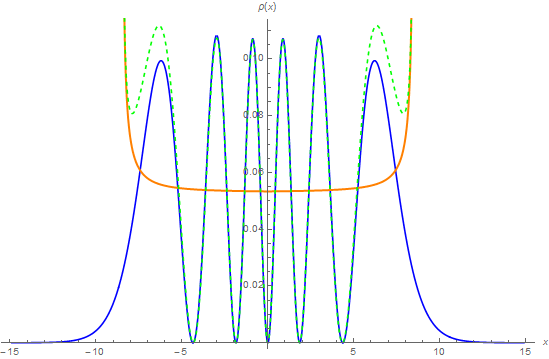}
\includegraphics[width=5cm,height=3.30cm]{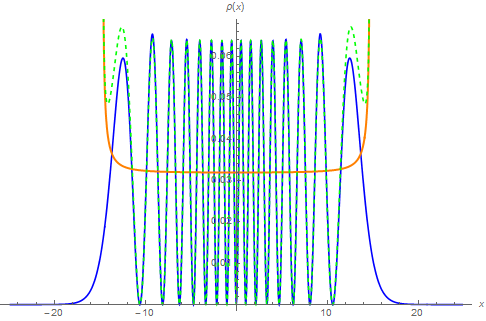}
\caption{Probability distributions $\rho_{\textrm{FGH}}(x)$ (solid blue), $\rho_{\textrm{WKBJ}}(x)$ (dashed green) and $\rho_{\textrm{cl}}(x)$ (bold solid orange) for Hamiltonian~(\ref{hsr}) with $m=\lambda=0.2$. From left to right: $n=0$, 5, 15.}
\label{fig1}
\end{figure}

In order to test the validity of the WKBJ approximation for a non-usual kinematics, eigenstates of the relativistic Hamiltonian, written in natural unit ($\hbar=c=1$),
\begin{equation}
\label{hsr}
H=\sqrt{p^2+m^2}+\lambda\,|x|
\end{equation}
have been computed. Such Hamiltonian (in 3D-space) are used to study hadrons in constituent quark models \cite{isgu85,buis12}. The numerical solutions are computed with the Fourier grid Hamiltonian (FGH) method \cite{mars89,sema00}, which is particularly well suited for the case of Hamiltonians with non-standard kinetic parts \cite{sema98,sema12}. The quantum probability distribution $\rho_{\textrm{FGH}}(x)$ obtained by this method has been compared with the corresponding distributions $\rho_{\textrm{WKBJ}}(x)$ and $\rho_{\textrm{cl}}(x)$. $\rho_{\textrm{FGH}}(x)$ is normalized to unity on $]-\infty,+\infty[$, while $\rho_{\textrm{WKBJ}}(x)$ and $\rho_{\textrm{cl}}(x)$ are normalized to unity on $]a,b[$. For such Hamiltonian, the integrals necessary for the computation of $\rho_{\textrm{WKBJ}}(x)$ are analytical, but they are not written here because of their complicated structure. It can be seen on Fig.~\ref{fig1} that the WKB approximation is quite good, even for values of $n$ as low as 5. With $m=\lambda=0.2$, the relative errors between the eigenvalues computed by the FGH method and the WKBJ approximation are respectively around $10^{-1}$,  $10^{-3}$, $10^{-4}$, for $n=0$, 5, 15. These results are similar for other computations performed with different finite values of the dimensionless ratio $m/\sqrt{\lambda}$. 

\begin{figure}[htb]
\includegraphics[width=5cm,height=3.26cm]{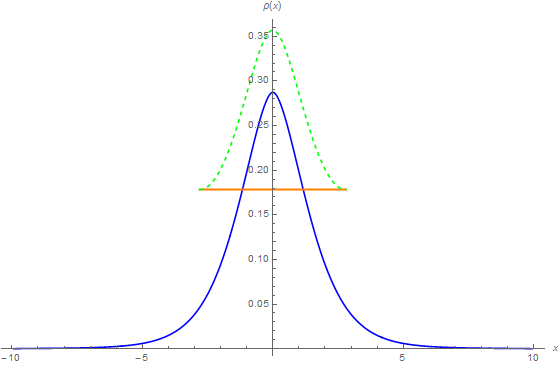}
\includegraphics[width=5cm,height=3.26cm]{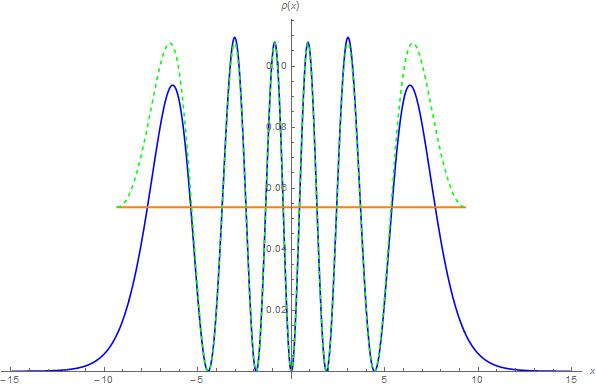}
\includegraphics[width=5cm,height=3.24cm]{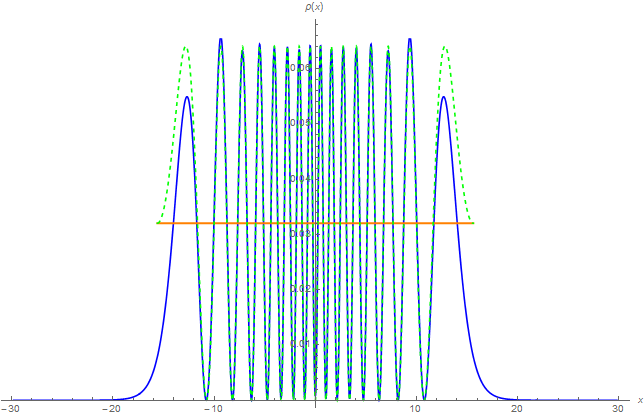}
\caption{Probability distributions $\rho_{\textrm{FGH}}(x)$ (solid blue), $\rho_{\textrm{WKBJ}}(x)$ (dashed green) and $\rho_{\textrm{cl}}(x)$ (bold solid orange) for Hamiltonian~(\ref{hsr}) with $m=0$ ($\lambda=0.2$ to fix the scale). From left to right: $n=0$, 5, 15.}
\label{fig2}
\end{figure}

The case $m=0$ in Hamiltonian~(\ref{hsr}) is particular since $T(p)=|p|$ is not derivable in $p=0$. Moreover, $|v(x)|=1$ and $\rho_{\textrm{cl}}(x)=1/d$. Some results are presented on Fig.~\ref{fig2}. Surprisingly, the WKBJ results are reasonable as well as the classical probability distribution, though expansion~(\ref{3.3a}) is not relevant. In this case, the relative errors between the eigenvalues computed by the FGH method and the WKBJ approximation are respectively around $2\times 10^{-1}$,  $3\times 10^{-3}$, $3\times 10^{-4}$, for $n=0$, 5, 15. This shows that this kind of approximation is probably quite robust. 

\section{Concluding remarks}
\label{sec:conclusion}

The WKBJ method \cite{flug99,maha09} yields a semi-classical solution of a one-dimensional Schr\"odinger equation. The approximate quantum probability distribution obtained is a good approximation of the genuine solution, in the limit of large quantum excitations. With an appropriate averaging procedure, this WKBJ distribution reduces to the classical probability distribution which can be defined for the corresponding classical systems \cite{robi95,yode06}. In this paper, all these results are generalized for one-dimensional Hamiltonians with an arbitrary kinetic energy. 

Only one-dimensional general Hamiltonians are considered here. But the results obtained can probably be generalized for systems living in spaces with more than one dimension, as it is the case of non-relativistic Hamiltonians \cite{pete87,sen05,mart13a,mart13b}.

\end{document}